\documentclass[12pt]{article}
\usepackage{amssymb}
\usepackage{color}

\newcommand{\R}{\Bbb R}
\newcommand{\Z}{\mathbb Z}
\newcommand{\mc}{\mathcal}

\newcommand{\be}{\begin{equation}}
\newcommand{\en}{\end{equation}}

\newcommand{\Lc}{{\cal L}}
\newcommand{\1}{1 \!\! 1}

\newcommand{\N}{\mathbb N}
\newcommand{\Hil}{\mc H}

\catcode `\@=11 \@addtoreset{equation}{section}

\catcode `\@=12

\begin{document}

\begin{center}
{\Large \bf Relations between multi-resolution analysis and  quantum mechanics}   \vspace{2cm}\\

{\large F. Bagarello}
\vspace{3mm}\\
  Dipartimento di Metodi e Modelli Matematici,
Facolt\`a di Ingegneria,\\ Universit\`a di Palermo, I - 90128  Palermo, Italy\\
E-mail: bagarell@unipa.it\\home page:
www.unipa.it$\backslash$\~\,bagarell
\vspace{2mm}\\
\end{center}

\vspace*{2cm}

\begin{abstract}
\noindent We  discuss a procedure to construct multi-resolution
analyses (MRA) of $\Lc^2(\R)$   starting from a given {\em seed}
function $h(s)$ which should satisfy some conditions.  Our method,
originally related to the quantum mechanical hamiltonian of the
fractional quantum Hall effect (FQHE), is shown to be model
independent. The role of a canonical map between certain
canonically conjugate operators is discussed. This clarifies our
previous  procedure and makes much easier most of the original
formulas, producing a convenient framework to produce examples of
MRA.

\end{abstract}

\vfill

\newpage

\section{Introduction}

Ia a series of papers, \cite{b1}-\cite{b7}, we have discussed the
relations between a generic MRA and the ground state of the free
single-electron hamiltonian of the FQHE. In particular we proved
that any MRA produces an orthonormal set of functions in the
subspace of $\Lc^2(\R^2)$ known as the {\em lowest Landau level},
see Section II, and that, vice-versa, any such a set produces a
sequence of complex numbers related to a certain MRA.

In this paper we extend these results and propose a {\em
model-independent} construction which still give rise to a MRA
starting from a certain square integrable function, which we call
{\em seed function}. Our extension clarifies the role of some
canonical maps for a certain quantum hamiltonian, and for its
related physical system, which is behind the construction.

\vspace{2mm}

 We devote the rest of this Introduction to recall,
just to fix the notation, few known facts about MRA which will be
useful in the following.

 A MRA of
$\Lc^2(\mathbb{R})$ is an increasing sequence of closed subspaces
\be \ldots \subset V_{-2} \subset V_{-1} \subset V_0 \subset V_1
\subset V_2\subset\ldots\subset\Lc^2(\mathbb{R}), \en with $
\bigcup_{j \in \mathbb{Z}} V_j$ dense in $\Lc^2(\mathbb{R})$ and $
\bigcap_{j \in \mathbb{Z}} V_j = \{0\}$, and such that
\begin{itemize}
\item[(1)] $f(x) \in  V_j \Leftrightarrow f(2x) \in  V_{j+1}$
\item[(2)] There exists a function $\phi \in V_0$, called {\em
scaling} function, such that  $\{\phi(x-k), k \in {\mathbb{Z}}\}$
is an o.n. basis of $V_0$.
\end{itemize}
From these two requirements clearly follows that, for any fixed
$j\in\mathbb{Z}$,  $ \{\phi_{j,k}(x) \equiv 2^{j/2} \phi(2^jx-k),$
$ k \in \mathbb{Z} \}$ is an o.n. basis of $V_j$, which can be
interpreted as an {approximation space}: the approximation of $f
\in \Lc^2(\mathbb{R})$ at the resolution $2^j$ is defined by its
projection onto $V_j$. The additional details needed for
increasing the resolution from $2^j$ to $2^{j+1}$ are given by the
projection of $f$ onto the orthogonal complement $W_j$ of $V_j$ in
$ V_{j+1}$: \be V_j \oplus W_j = V_{j+1}, \en and we have: \be
\bigoplus_{j \in {\mathbb{Z}}} W_j = \Lc^2(\mathbb{R}). \en

Now, the main result of a MRA is that there exists a function
$\psi$, {\em the mother wavelet}, explicitly computable from
$\phi$, such that $\{\psi_{j,k}(x) \equiv 2^{j/2} \psi(2^jx-k),
j,k \in \mathbb{Z}\}$ constitutes an orthonormal basis of $
\Lc^2(\mathbb{R})$.

The construction of  $\psi$ proceeds as follows. First, the
inclusion $V_0 \subset V_1$ yields the relation \be   \label{phi}
\phi(x) = \sqrt{2} \,\sum_{n=-\infty}^\infty \; h_n \phi(2x - n),
\quad h_n = \langle \phi_{1,n} | \phi \rangle. \en  Then one {
uses these coefficients} to define the function $\psi$ as \be
\label{psi} \psi(x) = \sqrt{2} \; \sum_{n=-\infty}^\infty
\;(-1)^{n-1} h_{-n-1}
              \phi(2x - n).
\en As we see, the role of the coefficients $h_n$ is quite
important. For this reason we introduce the following definition:

\vspace{2mm}

\noindent \underline{\bf Definition 1:}-- We call {\bf\em
 relevant} any sequence $h=\{h_n, n\in\mathbb{Z}\}$
which satisfies the following properties:

\begin{itemize}
\item[(r1)]\hspace{2cm}
$\sum_{n\in\mathbb{Z}}h_n \overline{h_{n+2l}}=\delta_{l,0}$;
\item[(r2)]\hspace{2cm}
$h_n=O(\frac{1}{1+|n|^2}), \quad n\gg 1;$
\item[(r3)]\hspace{2cm}
$\sum_{n\in\mathbb{Z}}h_n =\sqrt{2};$
\item[(r4)]\hspace{2cm}
$H(\omega)=\frac{1}{\sqrt{2}}\sum_{n\in\mathbb{Z}}h_ne^{-i\omega
n}\neq 0 \quad\quad \forall
\omega\in[-\frac{\pi}{2},\frac{\pi}{2}].$
\end{itemize}

Using Mallat's algorithm it is known that any relevant sequence
produces a MRA. In particular, it produces a scaling function
$\Phi(x)$ and the related mother wavelets, \cite{mal}.

The main goal of this paper is the construction of a quite
non-standard procedure which helps in the production of relevant
sequences and, as a consequence, of multi-resolutions of
$\Lc^2(\R)$. More in details:

in the next section we briefly resume our original results in this
direction related to the Hall effect.

In Section III we propose our more abstract approach, mainly
regarding  condition (r1) of Definition 1. This analysis will
produce a so-called {\em orthonormality condition}, ONC,  for a
certain {\em seed} function $h(s)$ in $\Lc^2(\R)$. We will show
how to use $h(s)$ to construct a sequence $\{h_n\}$ satisfying
condition (r1).

In Section IV we show how to find easily solutions of the ONC and,
as a consequence, how to produce sequences satisfying condition
(r1).

In Section V we propose an {\em orthonormalization trick}, ONT,
which generates more solutions of the ONC.

 In Section
VI we consider  the other requirements contained in Definition 1,
in connection with our approach.

Section VII contains our conclusions and plans for the future.

\section{The old results: FQHE}

The many-body model of the FQHE  consists simply in a
two-dimensional electron gas, 2DEG, (that is a gas of electrons
constrained in a two-dimensional layer) in a positive uniform
background and subjected to an uniform magnetic field along $z$,
whose hamiltonian (for $N$ electrons) is, \cite{bms},
\begin{equation}
H^{(N)}=H^{(N)}_0+\lambda(H^{(N)}_C+H^{(N)}_B),\label{31}
\end{equation}
where $H^{(N)}_0$ is the sum of $N$ contributions:
\begin{equation}
H^{(N)}_0=\sum^N_{i=1}H_0(i).\label{32}
\end{equation}
Here $H_0(i)$ describes the minimal coupling of the $i-$th
electron with the magnetic field:
\begin{equation}
H_0={1\over 2}\,\left(\underline p+\underline
A(r)\right)^2={1\over 2}\,\left(p_x-{y\over 2}\right)^2+{1\over
2}\,\left(p_y+{x\over 2}\right)^2. \label{33}
\end{equation}
$H^{(N)}_C$ is the canonical Coulomb interaction between charged
particles, $ H^{(N)}_C={1\over2}\,\sum^N_{i\not=j}{1
\over|\underline r_i- \underline r_j|},$ and $H^{(N)}_B$ is the
interaction of the charges with the background,  \cite{bms}.

We now consider $\lambda(H^{(N)}_C+H^{(N)}_B)$ as a perturbation
of the free hamiltonian $H^{(N)}_0$, and we  look for eigenstates
of $H^{(N)}_0$ in the form of Slater determinants built up with
single electron wave functions.  The easiest way to approach this
problem consists in introducing the  new variables
  \be
\label{35}
  P'= p_x-y/2, \hspace{5mm}     Q'= p_y+x/2.
  \en
In terms of $P'$ and $Q'$ the single electron hamiltonian, $H_0$,
can be written as
 \be
\label{36}
  H_{0}=\frac{1}{2}(Q'^2 + P'^2).
  \en
The transformation (\ref{35}) can be seen as a part of a canonical
map from $(x,y,p_x,p_y)$ into $(Q,P,Q',P')$ where

   \be
\label{37}
   P= p_y-x/2, \hspace{5mm}
  Q= p_x+y/2.
   \en
  These operators  satisfy the following commutation relations:
  \be
\label{38}
 [Q,P] = [Q',P']=i, \quad  [Q,P']=[Q',P]=[Q,Q']=[P,P']=0.
  \en
 Using the results contained in \cite{mo}, it can be deduced   that  a wave function in the $(x,y)$-space
is
   related to its  $PP'$-expression by the formula
  \be
\label{39}
  \Psi(x,y)=\frac{e^{ixy/2}}{2\pi}\int_{-\infty}^{\infty}\,
  \int_{-\infty}^{\infty}e^{i(xP'+yP+PP')}\Psi(P,P')\,dP dP',
  \en
which can be easily inverted:
  \be
\label{325}
  \Psi(P,P')=\frac{e^{-iPP'}}{2\pi}\int_{-\infty}^{\infty}\,
  \int_{-\infty}^{\infty}e^{-i(xP'+yP+xy/2)}\Psi(x,y)\,dx dy.
  \en
  The usefulness of the $PP'$-representation stems from the expression
(\ref{36})
  of $H_0$. Indeed, in this representation, the single electron Schr\"{o}dinger equation
admits
  eigenvectors $\Psi(P,P')$ of $H_0$ of the form $\Psi(P,P')=f(P')h(P)$.
Thus
   the ground state of (\ref{36}) must have the form
 $f_0(P')h(P)$,  where
   \be     \label{310}
  f_0(P')= \pi^{-1/4} e^{-P'^2/2},
  \en
  while the function $h(P)$ is arbitrary, which manifests the degeneracy of
the {\em lowest Landau level}, LLL, i.e. the lowest eigenspace of
$H_0$. The explicit expression of $h(P)$ should be fixed by the
interaction, whose mean value should be minimized. With $f_0$ as
above, formula (\ref{39}) becomes \be
 \label{311}
  \Psi(x,y) = \frac{e^{ixy/2}}{\sqrt{2}\pi^{3/4}}
\int_{-\infty}^{\infty}\,e^{iyP}e^{-(x+P)^2/2}h(P)\,dP,
 \en
whose {\em inverse} is \be
h(P)=\frac{e^{-iPP'+P'^2/2}}{2\pi^{3/4}}\int_{-\infty}^{\infty}\,
  \int_{-\infty}^{\infty}e^{-i(xP'+yP+xy/2)}\Psi(x,y)\,dx dy
\label{326} \en

Let us now define the so-called magnetic translation operators
$T(\vec{a_i})$ for a square lattice with basis $\vec{a_1}=a(1,0)$,
$\vec{a_2}=a(0,1)$, $a^2=2\pi$, \cite{b4},  by
   \be
  T_1:=T(\vec{a_1})=e^{iaQ}, \quad T_2:=T(\vec{a_2})=e^{iaP}.
\label{312}
  \en
We see  that, due to (\ref{38}) and to the condition on the cell
of the lattice, $a^2=2\pi$, \be
[T(\vec{a_1}),T(\vec{a_2})]=[T(\vec{a_1}),
H_0]=[T(\vec{a_2}),H_0]=0. \label{314} \en

The action of the $T$'s on a generic function $f(x,y)\in
\Lc^2(\R^2)$ is the following: \be f_{m,n}(x,y):=T_1^mT_2^n
f(x,y)=(-1)^{mn}e^{i\frac{a}{2}(my-nx)}f(x+ma,y+na). \label{316}
\en This formula shows that, if for instance $f(x,y)$ is localized
around the origin, then $f_{m,n}(x,y)$ is localized around the
site $a(-m,-n)$ of the square lattice.

Now we have all the ingredients to construct the ground state of
$H^{(N)}_0$ mimicking the classical procedure. We simply start
from the single electron ground state of $H_0$ given in
(\ref{311}), $\Psi(x,y)$. Then we construct a set of copies
$\Psi_{m,n}(x,y)$ of $\Psi(x,y)$ as in (\ref{316}), with
$m,n\in\Z$. All these functions still belong to the LLL for any
choice of the function $h(P)$ due to (\ref{314}). $N$ of these
wave functions $\Psi_{m,n}(x,y)$ are finally used to construct a
Slater determinant $\Psi^{(N)}$ for the finite system in the usual
way, which is normalized for all $N$ if \be
<\Psi_{m_i,n_i}\Psi_{m_j,n_j}>=\delta_{m_i,m_j}\delta_{n_i,n_j}.
\label{317bis} \en

Let $\Psi(x,y)$ be as in (\ref{311}) and $\Psi_{m,n}(x,y) =
(-1)^{mn} e^{i\frac{a}{2}(my-nx)} \Psi(x+ma,y+na). $  With the
above definitions we find \be \tilde
S_{l_1,l_2}=<\Psi_{0,0},\Psi_{l_1,l_2}>=\int_{-\infty}^\infty dp
e^{-il_2ap}\overline{h(p-l_1a)}h(p), \label{320} \en which
restates the problem of the orthonormality of the wave functions
in the LLL in terms of the unknown function $h(P)$.

In the construction above we are considering a square lattice in
which all the lattice sites are occupied by an electron. We say
that the {\em filling factor} $\nu$ is equal to 1. We have seen in
\cite{b4} that, in order to construct an o.n. set of functions in
the LLL corresponding to a filling $\nu=\frac{1}{2}$ (only half of
the lattice sites are occupied), we have to replace (\ref{320})
 with the following slightly weaker condition, $$
S_{l_1,l_2}=\tilde S_{l_1,2l_2}=\int_{-\infty}^\infty dp
e^{-2il_2ap}\overline{h(p-l_1a)}h(p)=$$ \be=\int_{-\infty}^\infty
dp e^{il_1ap}\overline{\hat h(p-2l_2a)}\hat
h(p)=\delta_{l_1,0}\delta_{l_2,0}, \label{321} \en for all
$l_1,l_2\in\Z$, where $\hat
h(p)=\frac{1}{\sqrt{2\pi}}\int_{\R}e^{-ipx}h(x)dx$ is the Fourier
transform of $h(x)$. If $h(x)$ satisfies (\ref{321}), then,
defining \be h_n=\frac{1}{\sqrt{a}}\int_{-\infty}^\infty dp
e^{-inxa}h(x), \label{322} \en it is easily checked that \be
\sum_{n\in\Z}h_n \overline{h_{n+2l}}=\delta_{l,0}. \label{323} \en
The proof of this claim, contained in \cite{b4}, is based on
condition (\ref{321}) and on the use of the Poisson summation
formula (PSF) which we write here as \be
\sum_{n\in\Z}e^{inxc}=\frac{2\pi}{|c|}\sum_{n\in\Z}\delta(x-n\frac{2\pi}{c}),
\label{324} \en for any $c\in\R$. It is well known that the PSF
does not always hold, see \cite{chen} p.298  and references
therein, for instance. In this paper, however, we will always
assume its validity.

In \cite{b7} we have also discussed a possible way to find
solutions of the equation (\ref{321}) starting from a generic seed
function in $\Lc^2(\R)$: in particular we have shown how the
assumption that this function produces a relevant sequence of
complex numbers, i.e. a sequence satisfying Definition 1 above,
produces many constraints on the seed function itself. We will now
go back to this construction from a much more abstract point of
view, showing that there exists a general framework behind the
construction just sketched, construction which allows us to
extract the really crucial ingredients of our method.

\vspace{2mm}

\underline{\bf Remark:} we also want to remind that the above
construction has been extended to other shapes of the lattice and
to different values of the filling. These results, contained in
\cite{b5}, are more relevant for concrete numerical applications
to the FQHE, but exactly for this same reason, are harder to be
concretely applied.

\section{A more abstract point of view}

In this section we will {\em embed} the above results in an
abstract and more general framework. This will make our procedure
more direct and much simpler, both from a theoretical and from a
practical point of view.

\vspace{4mm}

Consider the operators $((\hat x, \hat p_x),(\hat y, \hat p_y))$
and $((\hat x_1, \hat p_1),(\hat x_2, \hat p_2))$, satisfying
$$[\hat x,\hat p_x]=[\hat y,\hat p_y]=i,\hspace{1cm}[\hat x_1,\hat p_1]=[\hat x_2,\hat p_2]=i$$
Let $\xi_x$ and $\eta_y$ be the generalized eigenstates of $\hat
x$ and $\hat y$: $\hat x\xi_x=x\xi_x$, $\hat y\eta_y=y\eta_y$, and
$\xi'_{x_1}$ and $\eta'_{x_2}$ the eigenstates of  $\hat x_1$ and
$\hat x_2$: $\hat x_1\xi'_{x_1}=x_1\xi'_{x_1}$, $\hat
x_2\eta'_{x_2}=x_2\eta'_{x_2}$. We recall that all these vectors
are $\delta$-like normalized, e.g.,
$<\xi_x,\xi_{x'}>=\delta(x-x')$,
$<\eta_y,\eta_{y'}>=\delta(y-y')$, and  produce resolutions of the
identity: \be
\int\,dx\int\,dy\,|\xi_{x,y}><\xi_{x,y}|=\int\,dx_1\int\,dx_2\,|\xi'_{x_1,x_2}><\xi'_{x_1,x_2}|=\1,
\label{new1}\en where $\xi_{x,y}=\xi_{x}\otimes\eta_{y}$ and
$\xi'_{s,t}=\xi'_{s}\otimes\eta'_{t}$. In this section sometime we
adopt the Dirac {\em bra-ket} symbols  to simplify the notation.
Any $\Psi\in\Hil$, our Hilbert space, can be written in the
$(x,y)$-coordinates or in the $(x_1,x_2)$-coordinates as
$$\Psi(x,y)=<\xi_{x,y}|\Psi> \hspace{1cm}\mbox{and}\hspace{1cm}
\Psi'(x_1,x_2)=<\xi'_{x_1,x_2}|\Psi>,$$ which, because of
(\ref{new1}), are related to each other as follows
\be\Psi(x,y)=\int\,dx_1\int\,dx_2\,<\xi_{x,y}|\xi'_{x_1,x_2}>\Psi'(x_1,x_2)\label{new2}\en
and
\be\Psi'(x_1,x_2)=\int\,dx\int\,dy\,<\xi'_{x_1,x_2}|\xi_{x,y}>\Psi(x,y)\label{new3}\en

It is clear that these formulas are just the abstract versions of
formulas (\ref{39}) and (\ref{325}), with a kernel \be
K(x,y;x_1,x_2):=<\xi_{x,y}|\xi'_{x_1,x_2}>\label{new3bis}\en which
is easily identified.

We may interpret $(x,y)$ as the {\em physical} spatial coordinates
(in analogy with the FQHE), while $(x_1,x_2)$ can be seen as a
pair of fictitious coordinates and they are not required to have
any physical meaning, in general. For this reason there is no
objection in taking $\Psi'(x_1,x_2)$ as a product function
$\Psi'(x_1,x_2)=\varphi(x_1)h(x_2)$ in (\ref{new2}), and we call
$\Psi^{(h)}(x,y)$ the related function in the $(x,y)$-space: \be
\Psi^{(h)}(x,y)=\int\,dx_1\int\,dx_2\,K(x,y;x_1,x_2)\varphi(x_1)h(x_2)\label{new4}\en
As a matter of fact, $\Psi^{(h)}(x,y)$ clearly also depends on
$\varphi$. However, it will appear clear in the following that
this dependence disappears in all the scalar products we will
consider. For this reason we prefer to adopt this simpler but
somehow misleading notation.

 Also, we introduce three commuting operators: $H= H(\hat
x_1,\hat p_1)=H^\dagger$, $T_1=e^{ia\hat x_2}$ and $T_2=e^{ia\hat
p_2}$. Here, for reasons that will appear clear in the following,
we take $a^2=4\pi$\footnote{this is slightly different from what
we have done in the previous section, where we had $a^2=2\pi$ but
where only one site of the lattice every two was {\em occupied} by
an electron}. It is clear that, as for the FQHE, independently of
the explicit definition of $H$, we have\be
[T_1,T_2]=[T_1,H]=[T_2,H]=0.\label{new4bis}\en We still call the
unitary operators {\em magnetic translations} and $H$ the {\em
hamiltonian}. Notice that, while the explicit expressions for
$T_1$ and $T_2$ are fixed above, there is no need to fix the
expression of $H$, which will be kept general here. We will
comment on possible explicit expressions of $H$ several times
along the paper and in the examples below. What this really means
is that in our treatment there is no need of having any {\em
concrete} physical system behind.

As for the FQHE, the main idea is to require orthonormality of the
functions \be\Psi^{(h)}_{\vec
l}(x,y)=T_1^{\,l_1}T_2^{\,l_2}\Psi^{(h)}(x,y)=\int dx_1\int
dx_2\,K_{\vec l\,}(x,y;x_1,x_2) \varphi(x_1)h(x_2),\label{new5}\en
where $\vec l=(l_1,l_2)$ and $K_{\vec
l\,}(x,y;x_1,x_2)=T_1^{\,l_1}T_2^{\,l_2}K(x,y;x_1,x_2)$, which,
all together, {\em generate} a (fictitious) lattice with cell area
equal to $4\pi$. In particular, if the function $\varphi(x_1)$ is
an eigenstate of $H(\hat x_1,\hat p_1)$ corresponding to an
eigenvalue $\epsilon$, then  each $\Psi^{(h)}_{\vec l}(x,y)$ is
still an eigenstate of $H(\hat x_1,\hat p_1)$ with the same
eigenvalue $\epsilon$. We can still speak of {\em infinite
degeneracy} of the energetic levels, which we still call {\em
Landau levels}. In this case we could think of $H$ as an operator
like $\epsilon|\varphi><\varphi|+\tilde H$, where $\tilde H$ is
again self adjoint, and contains the rest of the spectrum of $H$
rather than $\epsilon$.

\vspace*{2mm}

\underline{\bf Remarks:-} (1) In the FQHE the function
$\varphi(x_1)$ was taken to be the ground state of $H=H_0$. In the
rest of this section we will show that this is quite unessential.

(2) It may be worthwhile to notice that the appearance of two
commuting unitary operators like $T_1$ and $T_2$ strongly suggests
the relevance of the $(k,q)$-representation behind our strategy.
This is not surprising since the $(k,q)$-representation was
exactly our starting point in our first approach to the problem of
finding an orthonormal set in the LLL. This has been originally
discussed in \cite{bms} and, more in connection with MRA, in
\cite{b4} and \cite{b6}. However, how it will be clear from our
treatment, our main results can be found without any use of this
representation.

\vspace{2mm}

As for the FQHE we now compute the overlap between different wave
functions, which can be written as follows: \be
S^{(h)}_{l_1,l_2}=<\Psi_{l_1,l_2}^{(h)},\Psi_{0,0}^{(h)}>=\int_{\R^2}dt\,dt'\,\overline{h(t)}\,\Gamma_{\vec
l\,}(t,t')\,h(t'),\label{new6} \en where we have introduced the
following quantities \be \Gamma_{\vec
l\,}(t,t')=\int_{\R^2}ds\,ds'\,\overline{\varphi(s)}\,Q_{\vec
l\,}(s,t;s',t')\varphi(s') \label{new7}\en and \be Q_{\vec
l\,}(s,t;s',t')=\int_{\R^2}dx\,dy\,\overline{K_{\vec
l\,}(x,y;s,t)}\,K(x,y;s',t').\label{new8}\en

It is now a simple exercise in quantum mechanics to compute
$Q_{\vec l}$ and $\Gamma_{\vec l}$. Indeed, since $K_{\vec
l\,}(x,y;s,t)=T_1^{\,l_1}T_2^{\,l_2}<\xi_{x,y}|\xi'_{s,t}>=
<\xi_{x,y}|T_1^{\,l_1}T_2^{\,l_2}\xi'_{s,t}>$, the first
resolution of the identity in (\ref{new1}) and the other
properties of the vectors $\xi'_{s,t}$ imply that
$$Q_{\vec
l\,}(s,t;s',t')=<T_1^{\,l_1}T_2^{\,l_2}\xi'_{s,t},\xi'_{s',t'}>=<e^{iatl_1}\xi'_{s,t-al_2},\xi'_{s',t'}>=$$
\be=e^{-iatl_1}\delta(s-s')\delta(t-al_2-t')\label{new9}\en It is
now evident that, putting this in (\ref{new7}), we get, for any
normalized $\varphi$, \be \Gamma_{\vec
l\,}(t,t')=e^{-iatl_1}\delta(t-al_2-t'),\label{new10}\en which, in
turns, produces \be S^{(h)}_{l_1,l_2}=\int_{\R}\,ds\, h(s)\,
\overline{h(s+al_2)}\,e^{-isal_1}=\int_{\R}\,dp\,\hat h(p)\,
\overline{\hat h(p-al_1)}\,e^{-ipal_2},\label{new11}\en writing
the result also in terms of the Fourier transform $\hat h(p)$ of
$h(s)$.

\vspace*{2mm}

\underline{\bf Remarks:-} (1) This formula clearly shows what we
have stated before: the overlap between differently localized wave
functions constructed as shown above is independent of the
particular physical model we may consider, as well as from the
details of the canonical transformation mapping $((\hat x, \hat
p_x),(\hat y, \hat p_y))$ into $((\hat x_1, \hat p_1),(\hat x_2,
\hat p_2))$. Also, it does not depend on the explicit expression
for $\varphi(x_1)$, as far it is normalized in $\Lc^2(\R)$. This
last result was already noticed in \cite{b4}, where we proved that
the ONC obtained from wave functions in the higher Landau levels
coincides with (\ref{321}).

(2) It is also worthwhile to stress that the reason why we have
chosen here $a^2=4\pi$, instead of $a^2=2\pi$ as in the previous
section, is to avoid a unnatural asymmetry between the indices
$l_1$ and $l_2$ (i.e., between the two orthogonal directions of
the lattice), which is present, e.g., in formula (\ref{321}) but
not in our new approach, see (\ref{new11}). We recall that, see
\cite{b7} and references therein, the factor 2 appearing in
(\ref{321}) has a double meaning: from one side, it refers to the
value $\frac{1}{2}$ of the filling factor $\nu$ for the electron
gas. From the other side, it corresponds to a 2-MRA, that is to a
MRA with dilation parameter equal to 2.  In \cite{b5} we have
extended the procedure to a filling $\nu=\frac{1}{d}$ or,
equivalently, to a d-MRA, $d\in\N$. The same extension can be
performed here: indeed it would be sufficient to choose
$a=\sqrt{2\pi\,d}$.

(3) We finally want to stress that the differences arising between
(\ref{321}) and (\ref{new11}) are not only a consequence of the
different choices of the value of $a^2$ in the two sections, but
also follow from a  different choice of the variables used to
describe the wave function {\em after the unitary transformation}.
Indeed, while in this section we have used as new coordinates the
eigenvalues of the new position operators $\hat x_1$ and $\hat
x_2$, in Section II, adopting the same choice made in
\cite{b1}-\cite{bms} as well as in the paper where this canonical
map was used for the first time in connection with the FQHE,
\cite{dz}, we have used as new coordinates the eigenvalues of the
new momenta operators, but for a minus sign. We will come back on
this point in Example 1 below, where we make uniform the notation.

\vspace{3mm}

Before going on with the relations of our procedure with relevant
sequences, we briefly discuss three examples of the above
construction.

\underline{\bf Example 1.}

As a first example we consider the FQHE already discussed in many
details in the previous section. In order to  uniform the
notation, we rewrite the results using the approach discussed in
this section. In particular we take $a^2=4\pi$ and we use
$\Psi'(x_1,x_2)$, see (\ref{new3}), instead of $\Psi(P,P')$, see
(\ref{325}).

With this in mind we notice that the kernel of the transformation,
which is slightly different from the one deduced by (\ref{39}), is
$$K(x,y;s,t)=\frac{1}{2\pi}\exp\left\{i(xt+ys-st-\frac{xy}{2})\right\}$$
Using now the analogous of (\ref{316}) and condition $a^2=4\pi$ we
deduce that
$$K_{\vec
l\,}(x,y;s,t)=\frac{1}{2\pi}\exp\left\{i((x+l_1a)t+(y+l_2a)s-st-\frac{xy}{2}-ixl_2a)\right\}
$$
It is now easy to check that $$Q_{\vec
l\,}(s,t;s',t')=\frac{1}{(2\pi)^2}\int_{\R^2}dx\,dy\,e^{-i((x+l_1a)t+(y+l_2a)s-st-\frac{xy}{2}-ixl_2a)}
\,e^{i(xt'+ys'-s't'-\frac{xy}{2})}=
$$
$$=e^{-iatl_1}\delta(s-s')\delta(t-al_2-t'),$$
exactly as in (\ref{new9}). The results for $\Gamma_{\vec l}$ and
$S^{(h)}_{\vec l}$ are direct consequences of this one, and
coincide with (\ref{new10}) and (\ref{new11}) respectively.

\vspace*{3mm}

\underline{\bf Example 2.}

The second example was originally introduced in \cite{b3}, as a
prototype of the FQHE in which the single electron hamiltonian and
the canonical transformations are different, and simpler, than
those considered in Section II. In particular we have
$H=\frac{1}{2}(\hat p_x^2+\hat x^2)+\frac{1}{2}\hat p_y^2+\hat
p_x\hat p_y$. The new variables are defined as $\hat x_1=\hat
p_x+\hat p_y$, $\hat p_1=-\hat x$, $\hat x_2=\hat p_y$ and $\hat
p_2=\hat x-\hat y$. Then we have $[\hat x_j,\hat p_j]=i$, for
$j=1,2$, while all the other commutators among these new operators
are zero. In terms of these operators the hamiltonian looks like
$H=\frac{1}{2}(\hat x_1^2+\hat p_1^2)$, whose ground state is
$\varphi(x_1)=\frac{1}{\pi^{1/4}}e^{-x_1^2/2}$. We also introduce
the unitary operators $T_1=e^{ia\hat x_2}$ and $T_2=e^{ia\hat
p_2}$, with $a^2=4\pi$. These operators commute between themselves
and with $H$, and act on a generic function $f(x,y)\in\Lc^2(\R^2)$
as follows:
$T_1^{\,l_1}T_2^{\,l_2}f(x,y)=e^{ial_2(x-y)}f(x,y+al_1)$.
Following the procedure discussed in \cite{mo}, we find that the
kernel of the transformation is
$$K(x,y;s,t)=\frac{1}{2\pi}\exp\left\{ix(s-t)+iyt\right\},$$ so
that $K_{\vec
l\,}(x,y;s,t)=\frac{1}{2\pi}\exp\left\{ial_2(x-y)+ix(s-t)+i(y+al_1)t\right\}$.
Therefore $$Q_{\vec
l\,}(s,t;s',t')=\frac{1}{(2\pi)^2}\int_{\R^2}dx\,dy\,e^{-ial_2(x-y)-ix(s-t)-i(y+al_1)t}e^{ix(s'-t')+iyt'}=
$$
$$=e^{-iatl_1}\delta(s-s')\delta(t-al_2-t'),$$
again as in (\ref{new9}). The results for $\Gamma_{\vec l}$ and
$S^{(h)}_{\vec l}$  directly follow, and clearly coincide with the
ones in (\ref{new10}) and (\ref{new11}).

\vspace*{3mm}

\underline{\bf Example 3.}

We now consider a third example which differs from the previous
ones since there is no hamiltonian structure behind. We consider
the following canonical transformation: $((\hat x,\hat p_x),(\hat
y,\hat p_y))$ $\longrightarrow$ $((\hat x_1,\hat p_1),(\hat
x_2,\hat p_2))$, where $\hat x_1=\hat x-\hat p_x$, $\hat p_1=\hat
p_x$, $\hat x_2=\hat p_y$ and $\hat p_2=-\hat y+\hat p_y$. Then
$[\hat x_j,\hat p_j]=i$, for $j=1,2$, while all the other
commutators among the new operators are zero. Following \cite{mo}
we deduce the expression for the kernel of the transformation:
$$K(x,y;s,t)=\frac{1}{2\pi}\exp\left\{\frac{i}{2}\left(x^2+2(yt-xs)+s^2-t^2\right)\right\},$$ so
that, since
$T_1^{\,l_1}T_2^{\,l_2}f(x,y)=e^{-ial_2y}f(x,y+al_1+al_2)$ and
  $a^2=4\pi$, we get $$K_{\vec
l\,}(x,y;s,t)=\frac{e^{-ial_2y}}{2\pi}\exp\left\{\frac{i}{2}\left(x^2+2((y-al_1+al_2)t-xs)+s^2-t^2\right)\right\}.$$
It is now a trivial exercise to check that, again, formulas
(\ref{new9}), (\ref{new10}) and (\ref{new11}) are recovered.

\vspace{4mm}

It is now quite easy to check that any MRA produces solutions of
the following {\em orthonormality condition} (ONC): \be
S^{(h)}_{l_1,l_2}=\delta_{l_1,0}\delta_{l_2,0}\label{new12}\en
Indeed,  any MRA is related to a relevant sequence $h=\{h_n,
n\in\Z\}$ satisfying, among the others, property (r1) of
Definition 1: $\sum_{n\in\Z}h_n\overline{h_{n+2l}}=\delta_{l,0}$.
Modifying a little bit a result already contained in \cite{b7}, we
define now
\begin{eqnarray}
 h_2(s)=\left\{
    \begin{array}{ll}
        \frac{1}{\sqrt{a}}\sum_{n\in\mathbb{Z}}h_ne^{-isna/2},\hspace{1cm} s\in[0,a[, \\
        0 \hspace{4.5cm} \mbox{otherwise}\\
       \end{array}
        \right.
        \label{new13}
 \end{eqnarray}
Then we deduce that $S^{(h)}_{\vec
l}=\int_{\R}h_2(s)\overline{h_2(s+al_2)}e^{-isal_1}=\delta_{l_2,0}\int_{\R}|h_2(s)|^2e^{-isal_1}$,
because of the support of $h_2$, and we find $S^{(h)}_{\vec
l}=\delta_{l_2,0}\sum_{m\in\Z}h_m\overline{h_{m+2l_1}}=\delta_{\vec
l,\vec 0}$. Therefore $h_2(s)$ is a solution of the ONC above.

However, our main interest here is to proceed in the opposite
direction: given a solution of the ONC (\ref{new12}) we would like
to obtain a relevant sequence. This is also the only relevant
point here, since we are in a more general settings than in
\cite{b7}, and there is no physical system behind our
construction, in general. For this reason, it may be of no
interest at all to obtain an o.n. basis in a fictitious LLL.

The way in which the elements of our tentative relevant sequence
should be defined is suggested by formula (\ref{new13}): if we
want to recover $h_n$ from $h_2(s)$ we have to compute the
following integral:
$\frac{1}{\sqrt{a}}\int_{\R}h_2(s)\,e^{isna/2}\,ds$. This suggests
to take exactly this formula as our definition of $h_n$, given a
generic $h(s)$, solution of the ONC. Therefore we put \be
h_n=\frac{1}{\sqrt{a}}\int_{\R}h(s)\,e^{isna/2}\,ds.\label{new14}\en
It is now with a simple application of the PSF that we can prove
that the sequence $\{h_n\}$ satisfies condition (r1). Indeed we
have
$$\sum_{n\in\Z}h_n\overline{h_{n+2l}}=\frac{1}{a}\sum_{n\in\Z}\int_{\R}h(s)\,e^{isna/2}\,ds
\int_{\R}\overline{h(t)}\,e^{-ita(n/2+l)}\,dt=$$
$$=\frac{1}{a}\int_{\R^2}ds\,dt
\,h(s)\overline{h(t)}\,e^{-ital}\sum_{n\in\Z}e^{ina/2(s-t)}=$$
$$=\sum_{n\in\Z}\int_{\R^2}ds\,dt
\,h(s)\overline{h(t)}\,e^{-ital}\delta(s-t-na)=
$$
$$=\sum_{n\in\Z}\int_{\R}ds
\,h(s)\overline{h(s-na)}\,e^{-isal}=\sum_{n\in\Z}\delta_{n,0}\delta_{l,0}=\delta_{l,0},$$
which is what we had to prove.

\vspace{2mm}

Before considering examples of our construction it is interesting
to consider some points:

the first one is the following: the plus sign in the exponential
in the definition of $h_n$, formula (\ref{new14}), could be
replaced by a minus sign; indeed the sequence that we obtain still
satisfies condition (r1). This can be proved explicitly or simply
noticing that, if $\tilde h_n=h_{-n}$, then, introducing $m=-n$,
$$ \sum_{n\in\Z}\tilde h_n\overline{\tilde
h_{n+2l}}=\sum_{m\in\Z}\tilde h_{-m}\overline{\tilde
h_{-m+2l}}=\sum_{m\in\Z} h_{m}\overline{
h_{m-2l}}=\delta_{-l,0}=\delta_{l,0}.
$$
The second remark is obvious: using $\hat h(p)$ we can simply
write equation (\ref{new14}) as \be h_n=\sqrt{\frac{2\pi}{a}}\hat
h\left(-\frac{an}{2}\right).\label{new15}\en Finally,  formula
(\ref{new11}) for $S^{(h)}_{\vec l}$ has an interesting and useful
consequence, as far as solutions of the ONC are concerned. It is
clear indeed that if $h(s)$ solves the ONC, then another solution
of the ONC is a function $m(s)$ which is the inverse Fourier
transform of a function $\hat m(p):=h(p)$. This is a trivial
consequence of the expression of $S^{(h)}_{\vec l}$ given in terms
of $h(s)$ and $\hat h(p)$. It is also clear that $h(s)$ and $m(s)$
produce {\bf different} relevant sequences!

\section{Solutions of the ONC and consequences}

We already noticed that any MRA produces a solution of the ONC
(\ref{new12}) as in (\ref{new13}). These are not the only
solutions of the ONC. Different solutions are given in the table
below, where we list  the solution of the ONC in terms of $h(s)$
or of its Fourier transform $\hat h(p)$ and the coefficients that
we find using (\ref{new14}) or (\ref{new15}).

\vspace*{3mm}

\noindent\begin{footnotesize}
 \hspace{-0.2cm}\begin{tabular}{||c||c||c||}
\hline \hline $h(s)$           &$\hat h(p)$ &$h_n$ \\
\hline \hline$\left\{
    \begin{array}{ll}
        \frac{1}{\sqrt{a}},\hspace{.5cm} s\in[0,a[,\\
        0 \hspace{1.0cm} \mbox{otherwise}\\
       \end{array}
        \right.$&
&$h_n=\delta_{n,0}$  \\
\hline \hline$\left\{
    \begin{array}{ll}
        \frac{e^{-ia}}{\sqrt{a}},\hspace{.5cm} s\in[0,a[,\\
        0 \hspace{1.0cm} \mbox{otherwise}\\
       \end{array}
        \right.$&
&$h_n=i(1-e^{-ia})\,\frac{1}{2\pi n-a}$  \\
\hline \hline$\left\{
    \begin{array}{ll}
       {\sqrt{\frac{2}{a}}},\hspace{.5cm} s\in[0,a/2[,\\
        0 \hspace{1.0cm} \mbox{otherwise}\\
       \end{array}
        \right.$&
& $\left\{\begin{array}{ll}
       h_0=\frac{1}{\sqrt{2}}, h_{2n}=0\, \forall\, n\neq 0\\
        h_{2n+1}=\frac{i\sqrt{2}}{\pi(2n+1)}\\
       \end{array}
        \right.$  \\
\hline \hline$\left\{
    \begin{array}{ll}
       {\sqrt{\frac{2}{a}}},\hspace{.5cm} s\in[a/2,a[,\\
        0 \hspace{1.0cm} \mbox{otherwise}\\
       \end{array}
        \right.$&
&$\left\{\begin{array}{ll}
       h_0=\frac{1}{\sqrt{2}}, h_{2n}=0\, \forall\, n\neq 0\\
        h_{2n+1}=\frac{\sqrt{2}}{i\pi(2n+1)}\\
       \end{array}
        \right.$  \\
\hline \hline &$\left\{
    \begin{array}{ll}
        \frac{1}{\sqrt{a}},\hspace{.5cm} p\in[0,a[,\\
        0 \hspace{1.0cm} \mbox{otherwise}\\
       \end{array}
        \right.$&
$h_n=\frac{1}{\sqrt{2}}(\delta_{n,0}+\delta_{n,-1})$  \\
\hline \hline &$\left\{
    \begin{array}{ll}
        \sqrt{\frac{2}{a}},\hspace{.5cm} p\in[0,a[,\\
        0 \hspace{1.0cm} \mbox{otherwise}\\
       \end{array}
        \right.$&
$h_n=\delta_{n,0}$  \\
\hline \hline &$\left\{
    \begin{array}{ll}
       \frac{1}{\sqrt{2a}},\hspace{.5cm} p\in[0,a/2[,\cup[2a,3a[\\
       -\frac{1}{\sqrt{2a}},\hspace{.5cm} p\in[a/2,a[,\\
        0 \hspace{1.0cm} \mbox{otherwise}\\
       \end{array}
        \right.$
&$h_n=\frac{1}{2}(\delta_{n,0}+\delta_{n,-4}+\delta_{n,-5}-\delta_{n,-1})$  \\
\hline \hline &$\left\{
    \begin{array}{ll}
       \frac{1}{\sqrt{2a}},\hspace{.5cm} p\in[0,a/2[,\cup[a,2a[\\
       -\frac{1}{\sqrt{2a}},\hspace{.5cm} p\in[a/2,a[,\\
        0 \hspace{1.0cm} \mbox{otherwise}\\
       \end{array}
        \right.$
&$h_n=\frac{1}{2}(\delta_{n,0}+\delta_{n,-2}+\delta_{n,-3}-\delta_{n,-1})$  \\

\hline\hline
\end{tabular}
\end{footnotesize}

\vspace{3mm}

Moreover, it is also easy to check that  the  functions
$$h_{e1}(s)=\left\{
    \begin{array}{ll}
       \frac{1}{\sqrt{2a}},\hspace{.5cm} p\in[0,a/2[,\cup[2a,3a[\\
       -\frac{1}{\sqrt{2a}},\hspace{.5cm} p\in[a/2,a[,\\
        0 \hspace{1.0cm} \mbox{otherwise}\\
       \end{array}
        \right.
        \hspace{.4cm}
h(s)_{e2}=\left\{
    \begin{array}{ll}
       \frac{1}{\sqrt{2a}},\hspace{.5cm} p\in[0,a/2[,\cup[a,2a[\\
       -\frac{1}{\sqrt{2a}},\hspace{.5cm} p\in[a/2,a[,\\
        0 \hspace{1.0cm} \mbox{otherwise}\\
       \end{array}
        \right.
$$
both return the same coefficients as in the third row of the table
above: $h_0=\frac{1}{\sqrt{2}}$, $h_{2n}=0$ if $n\neq 0$, and
$h_{2n+1}=\frac{i\sqrt{2}}{\pi(2n+1)}$.

This is a first example of an interesting feature of our
procedure: the same sequence $\{h_n\}$ can be obtained starting
from very different functions $h(s)$. Just to mention another,
maybe more interesting, example of this fact let us consider the
Haar multiresolution. This can be obtained from the function
$$\hat h(p)=\left\{
    \begin{array}{ll}
        \frac{1}{\sqrt{a}},\hspace{.5cm} p\in[0,a[,\\
        0 \hspace{1.0cm} \mbox{otherwise},\\
       \end{array}
        \right.
 \,\mbox{ or by }\,\,  h_2(s)=\left\{
    \begin{array}{ll}
        \frac{1}{\sqrt{2a}}(1+e^{-isa/2}),\hspace{.5cm} s\in[0,a[,\\
        0 \hspace{1.0cm} \mbox{otherwise},\\
       \end{array}
        \right.$$ see the fifth row in the table above and formula (\ref{new13}).

 All the coefficients
$\{h_n\}$ found with our procedure satisfy condition (r1): this is
trivially checked in some of the examples above while it is
absolutely non trivial for the second, third and fourth examples
listed in the table. Also, it is worth noticing that the fifth
example produces the well known Haar MRA.

Moreover, we see from the first and the fifth row an example of
the symmetry that was mentioned at the end of the previous
section: we see that $h(s)$ and $\hat h(p)$ have the same
dependence on their variables but they produce different
coefficients. The same holds true for the seventh row and the
example arising from the function $h_{e1}(s)$ above: they have the
same dependence on, respectively, $p$ and $s$, but produce
completely different coefficients.

It is not hard to check that these coefficients do not always
generate relevant sequences: for instance, the sixth example of
the table give rise to a sequence which surely satisfies (r1) and
(r2), while condition (r3) does not hold. The fourth example, on
the contrary, satisfies all the conditions of Definition 1. We
will return on this point in Section VI.

\section{More solutions: the orthonormalization trick}

The solutions of the ONC considered in the previous section all
share a common feature: $h(s)$ or $\hat h(p)$ in the table are all
compactly supported. Moreover, most of the time, the support is
just contained in $[0,a[$. There is a reason for that: because of
the expression (\ref{new11}) of $S^{(h)}_{\vec l}$, these choices
produce $S^{(h)}_{\vec l}=\delta_{l_2,0}s_{l_1}$ or $S^{(h)}_{\vec
l}=\delta_{l_1,0}\tilde s_{l_2}$, with $s_{l_1}$ or $\tilde
s_{l_2}$ to be computed, depending on which function, $h(s)$ or
$\hat h(p)$, has compact support in $[0,a[$. One may wonder if
other solutions of the ONC do exist and, if they exist, how they
can be found.

In this section we will discuss an explicit construction which
allows, given a generic function $h(s)$ in $\Lc^2(\R)$, to
construct another function, $H(s)$, which is still in $\Lc^2(\R)$
and satisfies the ONC. We call this technique {\em
orthonormalization trick} (ONT) in analogy to what is done in
\cite{dau}, where it is shown how to use spline functions to
construct an orthonormal set in $V_0$.

Let $h(s)\in\Lc^2(\R)$ be a generic function,
$\Psi^{(h)}_{l_1,l_2}$ the related set in $\Lc^2(\R^2)$
constructed as discussed in Section III, and  $S^{(h)}_{l_1,l_2}=
<\Psi^{(h)}_{l_1,l_2},\Psi^{(h)}_{0,0}>$ the related overlap. If
$S^{(h)}_{l_1,l_2}\neq \delta_{l_1,0}\delta_{l_2,0}$ then the
$\Psi^{(h)}_{l_1,l_2}(\vec r)$ generate a non o.n. lattice (in
$\Lc^2(\R^2)$) and if we use $h(s)$, as in (\ref{new14}), to
construct a sequence $\{h_n\}$, this will not satisfy condition
(r1), in general. However, \cite{bms}, we can find an o.n. set in
$\Lc^2(\R^2)$, still invariant under magnetic translations, simply
by considering \be\Psi^{(H)}_{l_1,l_2}(\vec
r)=T_1^{l_1}T_2^{l_2}\Psi^{(H)}_{0,0}(\vec r), \mbox{ where
}\Psi^{(H)}_{0,0}(\vec r)=\sum_{\vec n\in\mathbb{Z}^2}f_{\vec
n}\,\Psi^{(h)}_{\vec n}(\vec r),\label{V1}\en and the coefficients
$f_{\vec n}$ are fixed by requiring that \be
S^{(H)}_{l_1,l_2}=\delta_{l_1,0}\delta_{l_2,0} \label{V2}\en In
these formulas a new function $H(s)$ has been implicitly
introduced:  $\Psi^{(H)}(\vec r)=\Psi^{(H)}_{0,0}(\vec r)$ is the
function in $\Lc^2(\R^2)$, (or in the LLL), which is generated via
formula (\ref{new4}) by the function $H(s)$. Again the role of the
function $\varphi$ is unessential, as far as the overlap between
the functions $\Psi^{(H)}_{\vec l}(\vec r)$ is concerned.  Without
giving the details of our construction, which are not really
different from those in \cite{b7}, defining the functions $F(\vec
p)=\sum_{\vec n\in\Z^2}f_{\vec n}e^{i\vec p\cdot\vec n}$ and
$S^{(h)}(\vec p)=\sum_{\vec n\in\Z^2}S^{(h)}_{\vec n}e^{i\vec
p\cdot\vec n}$, the orthonormality requirement (\ref{V2}) produces
$$\delta_{\vec l,\vec 0}=\sum_{\vec n,\vec m\in\Z^2}\overline{f_{\vec
n}}\,f_{\vec m}S^{(h)}_{\vec n+\vec l-\vec m}\Longrightarrow
1=|F(\vec p)|^2S^{(h)}(\vec p)\Longrightarrow F(\vec
p)=\frac{1}{\sqrt{S^{(h)}(\vec p)}},$$ where we have chosen
properly the phase in the solution (see \cite{b7} for a discussion
concerning the effects of the phase). The coefficients $f_{\vec
n}$ are therefore given by \be f_{\underline
s}=\frac{1}{(2\pi)^2}\int_0^{2\pi}\int_0^{2\pi}d^2\underline
p\frac{e^{-i\underline p\cdot \underline
s}}{\sqrt{S^{(h)}(\underline p)}}. \label{V3} \en and we find that
$H(s)=\sum_{\vec n\in\mathbb{Z}^2}f_{\vec
n}\,e^{isan_1}h(s+an_2)$.
 The related coefficients, $H_n=\frac{1}{\sqrt{a}}\int_{\R}\,ds\,H(s)\,e^{iasn/2}= \sqrt{\frac{2\pi}{a}}\,\hat
H\left(-\frac{na}{2}\right),$
 can be  finally written as
\be
H_n=\frac{1}{\sqrt{a}}\int_{\R}\frac{ds}{\sqrt{S^{(h)}(as,0)}}\,h(s)\,e^{iasn/2}.
\label{V4}\en

{\underline{\bf Remarks:}} (1) This formula should be compared
with equation (3.15) of \cite{b7}, which looks more difficult than
(\ref{V4}) to be concretely applied, since it involves an
(infinite) sum of integrals! It is clear, therefore, that
(\ref{V4}) represents a substantial improvement of our previous
results concerning the ONT.

(2) It is interesting to notice that, if from the very beginning
$S^{(h)}_{\vec
l}=<\Psi^{(h)}_{l_1,l_2},\Psi^{(h)}_{0,0}>=\delta_{l_1,0}\delta_{l_2,0}$,
then $S^{(h)}(\vec p)=1$ and we get $H_n=h_n$. In this case,
therefore, the  ONT does not modify the set of coefficients we get
from our procedure.

(3) It is a simple exercise to check that, using formula
(\ref{V4}), condition (r1) directly follows:
$\sum_{n\in\mathbb{Z}}H_n\overline{H_{n+2l}}=\delta_{l,0}$, for
all integer $l$. This result can be deduced by making use of the
PSF.

(4) As it can be deduced from a previous remark, it turns out that
it is sufficient to look to some {\em mild} version of the ONC
(\ref{new12}), like \be
S^{(h)}_{l_1,l_2}=\delta_{l_1}\,s_{l_2},\label{V5}\en for no
matter which $l^2(\Z)$-sequence $\{s_{l_2}\}$. We call this MONC.
We only need to require that this sequence belongs to $l^1(\Z)$,
as we will now see. Indeed if $h(s)$ solves (\ref{V5}) above, then
$S^{(h)}(p,0)=\sum_{l_1,l_2\in\mathbb{Z}}
S^{(h)}_{l_1,l_2}e^{ip\,l_1}=
\sum_{l_2\in\mathbb{Z}}s_{l_2}=:\sigma$, which is finite since
$\{s_{l_2}\}\in l^1(\Z)$ and does not depend on $p$. Therefore
equation (\ref{V4}) produces $H_n=\frac{h_n}{\sqrt{\sigma}}$, for
all $n\in\Z$. This result has the following meaning: if we have a
solution $h(s)$ of the MONC then its related coefficients $h_n$
also satisfy (r1) but for an over all normalization constant,
$\sigma$. In this case, therefore, the ONT reduces to a simple
overall normalization of the coefficients.







\section{On the other conditions}
In this section we briefly analyze the other conditions which make
of a sequence a {\em relevant one}. For all these conditions we
will discuss separately the cases in which $h(s)$ is a solution of
the ONC or not, so that the ONT is needed.

\subsection{On the asymptotic behavior}
Condition (r2) states that the asymptotic behavior of a relevant
sequence should be  as follows: $h_n=O(\frac{1}{1+|n|^2}), \, n\gg
1$. Suppose that $h_n$ are generated as in (\ref{new14}) or,
equivalently, as in (\ref{new15}), where $h(s)$ solves the ONC.
Then it is obvious that the requirement on the  asymptotic
behavior of $h_n$ is surely satisfied if, e.g., the function $\hat
h(p)$ has compact support. Suppose that this is not the case but
$h(s)$ is differentiable and $h'(s)$ still belongs to $\Lc^2(\R)$.
Then it is a standard exercise to check that there exists a
positive constant $M$ such that $|h_n|\leq \frac{M}{n}$ $\forall
n\in\Z$. It is clear then that the more regular $h(s)$ is, the
faster the coefficients $h_n$'s go to zero with $n$. In particular
then, in order (r2) to be satisfied, it is sufficient to look for
solutions of the ONC for which the second derivative exists and
still belongs to $\Lc^2(\R)$.

Suppose now that $h(s)$ does not satisfy the ONC. Then we need to
apply the ONT, which produces a new set of coefficients $H_n$ as
in (\ref{V4}). We can simply repeat  the above considerations
simply replacing $h(s)$ with $h(s)/\sqrt{S^{(h)}(as,0)}$, or with
its Fourier transform.

\vspace{2mm}

{\underline{\bf Remark:}} It may be worth noticing that these
results essentially simplify the ones obtained in \cite{b7},
where, in order to analyze the asymptotic behavior of the $h_n$'s,
we needed to use some non trivial results on the convolution of
sequences.

\subsection{Another sum rule}

We want now to comment briefly on condition (r3) of a relevant
sequence. As before, we first consider the case in which $h(s)$
solves the ONC. In this case, due to (\ref{new15}), it is clear
that (r3) is satisfied if and only if $\sum_{n\in\Z}\hat
h\left(\frac{na}{2}\right)=\sqrt{\frac{a}{\pi}}$. Using the PSF we
can also deduce that a necessary condition for (r3) to hold is
that $\sum_{n\in\mathbb{Z}}h(na)=\sqrt{\frac{2}{a}}$. This
condition is often deep in contrasts with condition (r1), how
appears clear from the examples discussed in Section IV, where
only one among the seven examples listed in the table satisfies
also this requirement. For this reason it is worth finding more
and more solutions of the ONC, in order to have a larger set of
possible solutions of condition (r3). For that the ONT is clearly
to be adopted. In this case, if we start with a generic
square-integrable function $h(s)$, we produce the set $H_n$ as in
(\ref{V4}), and we know that the set $\{H_n\}$ surely satisfies
condition (r1) for any given $h(s)$. However, in order for
$\{H_n\}$ to satisfy also condition (r3), not all the functions
$h(s)$ work equally well. In fact, it is possible to check that
the following must holds \be \sum_{n\in\Z} h(na)=\sqrt{\sum_{\vec
l\in\Z^2}h\left(\frac{al_1}{2}\right)\overline{h\left(\frac{al_1}{2}+al_2\right)}}
\label{VI1}\en or, in terms of the Fourier transform $\hat h(p)$
of $h(s)$, \be\sum_{n\in\mathbb{Z}}\hat
h\left(\frac{an}{2}\right)=\sqrt{2\,\sum_{\vec
l\in\mathbb{Z}^2}\hat h\left(\frac{al_2}{2}\right) \overline{\hat
h\left(\frac{al_2}{2}+al_21\right)}} \label{VI2}\en It is easy to
check, for instance, that the Haar example as given in Section IV
satisfies both these conditions.

To deduce condition (\ref{VI1}) we start observing that, by means
of the PSF,
$$\sqrt{2}= \sum_{n\in\Z}H_n=\frac{1}{\sqrt{a}} \sum_{n\in\Z}
\int_{\R}\frac{ds}{\sqrt{S^{(h)}(as,0)}}\,h(s)\,e^{iasn/2}=$$
$$=\frac{1}{\sqrt{a}} \sum_{n\in\Z}
\int_{\R}\frac{ds}{\sqrt{S^{(h)}(as,0)}}\,h(s)\,a\delta(s-an)=\sqrt{\frac{a}{S^{(h)}(\vec
0)}}\sum_{n\in\Z}h(na),$$ where we have also used that $a^2=4\pi$.
Condition (\ref{VI1}) follows now from an explicit computation of
$S^{(h)}(\vec 0)$, again performed by means of the PSF:
$$
S^{(h)}(\vec 0)=\sum_{\vec l\in\Z^2}S^{(h)}_{\vec l}=\sum_{\vec
l\in\Z^2}\int_{\R}\,ds\, h(s)\,
\overline{h(s+al_2)}\,e^{-isal_1}=$$ $$=\frac{a}{2}\sum_{\vec
l\in\Z^2}\int_{\R}\,ds\, h(s)\,
\overline{h(s+al_2)}\,\delta(s-\frac{a}{2}l_1)=\frac{a}{2}\sum_{\vec
l\in\Z^2}\left(\frac{al_1}{2}\right)\overline{h\left(\frac{al_1}{2}+al_2\right)}.$$

Equation (\ref{VI2}) simply follows from the definition of the
Fourier transform and, again, from the PSF.

\subsection{The last condition}

The last step consists is rephrasing condition (r4),
$h(\omega):=\frac{1}{\sqrt{2}}\sum_{n\in\mathbb{Z}}h_ne^{-i\omega
n}\neq 0$ for all $\omega\in\left[-\frac{\pi}{2},
\frac{\pi}{2}\right]$, in terms of the seed function. This can be
easily done and the condition we get turns out to be independent
of the fact that we need to use the ONT or not. Indeed, if $h(s)$
is already a solution of the ONC, then $h(\omega)$ can be
rewritten as
$$
h(\omega)=\frac{1}{\sqrt{2a}}\sum_{n\in\mathbb{Z}}\int_{\R}h(s)e^{ina/2(s-2\omega/a)}\,ds=
\sqrt{\frac{a}{2}}\sum_{n\in\mathbb{Z}}h\left(a\left(n+\frac{\omega}{2\pi}\right)\right),$$
using again the PSF. It is clear, then, that in order for (r4) to
hold, $h$ must satisfy the following condition: \be
\sum_{n\in\mathbb{Z}}h\left(a\left(n+\frac{\omega}{2\pi}\right)\right)\neq
0, \quad \omega\in\left[-\frac{\pi}{2},
\frac{\pi}{2}\right].\label{VI3}\en It is interesting to observe
that this same condition must be satisfied also when the seed
function does not solve the ONC, at least under very general
assumptions. Indeed,  in this case, condition (\ref{VI3}) should
be replaced by $\sum_{n\in\mathbb{Z}}\tilde
h\left(a\left(n+\frac{\omega}{2\pi}\right)\right)\neq 0,$ for all
$\omega\in\left[-\frac{\pi}{2}, \frac{\pi}{2}\right]$, where
$\tilde h(s)=\frac{h(s)}{\sqrt{S^{(h)}(as,0)}}$. But, since
$S^{(h)}(s,0)$ is $2\pi$-periodic, our requirement is satisfied
when (\ref{VI3}) holds, at least if $S^{(h)}(2\omega,0)$ is
bounded. This is always so in all the examples considered so far,
as well as any time $h(s)$ or $\hat h(p)$ are compactly supported.

It is worth remarking that this result simplify in a significant
way the original ones discussed in \cite{b7}.

\section{Conclusions}

We have discussed in some details a procedure to construct
relevant sequences, and MRA as a consequence, starting with a
given seed function $h(s)\in\Lc^2(\R)$ which should satisfy some
conditions. In particular, we have shown that no condition at all
is required to $h(s)$ as far as condition (r1) is concerned.
However, if $h(s)$ has to generate a sequence satisfying also
(r2)-(r4), not all the choices are equivalently good. Our method,
originally related to the quantum mechanical hamiltonian of the
FQHE, has been shown to be model-independent. What is really
relevant is the presence of a canonical map between certain
canonically conjugate operators. This makes all the procedure much
easier and suggests that many other applications might exist, for
example to the toy model discussed in \cite{b3}. Also, this map
allows us to construct a sort of quantum mechanical interpretation
of the ONT, relating this procedure to the construction of a
two-dimensional (fictitious) lattice of orthonormal functions of
$\Lc^2(\R^2)$, which can be chosen to belong all to a same
subspace of $\Lc^2(\R^2)$, the eigenspace of a certain {\em
hamiltonian} corresponding to a fixed eigenvalue.

Many other things still has to be done:

first of all we should construct more explicit examples of our
procedure. This will be done in a paper which is now in
preparation, \cite{b8}.

Secondly, it can be of some interest to deduce which conditions
should be imposed on our seed function in order to have stronger
conditions on the related mother wavelet.

Also, how it has been discussed in a rather recent paper by Ali
and myself, \cite{ab}, there exists an underlying modular
structure connected to the hamiltonian structure of the Hall
effect. This structure is still present in our present formulation
of the problem, and we believe that it deserves a deeper
investigation.

\section*{Acknowledgements}

This work has been financially supported in part by M.U.R.S.T.,
within the  project {\em Problemi Matematici Non Lineari di
Propagazione e Stabilit\`a nei Modelli del Continuo}, coordinated
by Prof. T. Ruggeri.

\end{document}